\documentclass[figures,preprint]{epl}   
\usepackage{amsfonts}
\usepackage{amssymb}
\usepackage{amsbsy}
\usepackage{graphics}
\usepackage{psfig}

\input epsf.sty

\newcommand{\BEQ}{\begin{equation}}     
\newcommand{\BEA}{\begin{eqnarray}}
\newcommand{\EEQ}{\end{equation}}       
\newcommand{\EEA}{\end{eqnarray}}
\def\be{\begin{equation}}
\def\ee{\end{equation}}
\def\ba{\begin{eqnarray}}
\def\ea{\end{eqnarray}}
\newcommand{\eps}{\varepsilon}          
\newcommand{\D}{{\rm d}}                

\renewcommand{\vec}[1]{\boldsymbol{#1}} 

\title{Scaling of the magnetic linear response in
phase-ordering kinetics}
\shorttitle{Scaling of the magnetic linear response}

\author{Malte Henkel\inst{1}, Matthias Pae{\ss}ens\inst{1,2}, and
Michel Pleimling\inst{3}}

\institute{
\inst{1}Laboratoire de Physique des Mat\'eriaux,
\footnote{Laboratoire associ\'e au CNRS UMR 7556}
Universit\'e Henri Poincar\'e Nancy I, B.P. 239,
F -- 54506 Vand{\oe}uvre l\`es Nancy Cedex, France\\
\inst{2} Institut f\"ur Festk\"orperforschung (Theorie II),
Forschungszentrum J\"ulich,
D -- 52425 J\"ulich, Germany\\
\inst{3} Institut f\"ur Theoretische Physik I, 
Universit\"at Erlangen-N\"urnberg,
D -- 91058 Erlangen, Germany}

\begin{document}
\maketitle

\begin{abstract}
The scaling of the thermoremanent magnetization and of the dissipative part
of the non-equilibrium magnetic susceptibility is analysed as a function of the
waiting-time $s$ for a simple ferromagnet undergoing phase-ordering kinetics 
after a quench into the ferromagnetically ordered phase. 
Their scaling forms describe the cross-over between two power-law
regimes governed by the non-equilibrium exponents $a$ and $\lambda_R/z$,
respectively. A relation between $a$, the dynamical
exponent $z$ and the equilibrium exponent $\eta$ is derived 
from scaling arguments. Explicit tests in the
Glauber-Ising model and the kinetic spherical model are presented. 
\end{abstract}

The study of non-equilibrium critical phenomena originated from
studies of the dynamic behaviour of glassy systems, but has also produced
new insights into the behaviour of the conceptually simpler ferromagnetic
systems, see \cite{Bouc00,Bray94,Godr02,Cugl02} for reviews. 
Since the latter are considerably more tractable, it might be hoped that
insights gained from studying the kinetics of simple ferromagnets could
provide clues for the comprehension of the former. In this letter, we 
consider simple ferromagnetic spin systems, quenched from a disordered
initial state to below its critical temperature $T_c>0$. We shall work
with a non-conserved order parameter throughout. 

Then the system undergoes phase ordering, that
is domains of a time-dependent typical size $L(t)\sim t^{1/z}$ form and
grow, where $z$ is the dynamical exponent \cite{Bray94}. The resulting slow
evolution of the system is more fully revealed through the
study of {\em two-time} quantities, such as the two-time autocorrelation
function $C(t,s)$ and the autoresponse function $R(t,s)$
\BEQ
C(t,s) = \langle \phi(t) \phi(s) \rangle \;\; , \;\;
R(t,s) = \left.\frac{\delta\langle\phi(t) \rangle}{\delta h(s)}\right|_{h=0}
\EEQ
where $\phi$ is the order parameter, $h$ the conjugate magnetic field,
$t$ is called the observation time and $s$ the waiting time. Ageing 
occurs in the regime when $s$ and $\tau=t-s>0$ are simultaneously much larger
than any microscopic time scale $\tau_{\rm micro}$. In many systems, one finds
in the ageing regime a scaling behaviour, see \cite{Bouc00,Godr02} 
\BEQ \label{gl:CR}
C(t,s) = s^{-b} f_C(t/s) \;\; , \;\; R(t,s) = s^{-1-a} f_R(t/s)
\EEQ
where $a$ and $b$ are non-equilibrium exponents and the scaling functions
behave for large arguments $x=t/s\gg 1$ asymptotically as
\BEQ
f_C(x) \sim x^{-\lambda_C/z} \;\; , \;\; f_R(x) \sim x^{-\lambda_R/z}
\EEQ
where $\lambda_C$ and $\lambda_R$ are the autocorrelation \cite{Fish88,Newm90} 
and autoresponse \cite{Pico02} exponents, respectively. 
For the usually studied case of an initial state without long-range 
correlations, it is generally accepted that $\lambda_C=\lambda_R=\lambda$. 
Combinations of rigorous results and of heuristic scaling arguments were used 
to derive the bounds $d/2\leq \lambda\lesssim d$ \cite{Fish88,Yeun96}. 

It has been proposed recently that one might be able to go beyond 
mere dynamical scale invariance as expressed 
by eq.~(\ref{gl:CR}) to a group of
{\em local} scale transformations related to conformal transformations in 
time \cite{Henk02}. If that hypothesis applies, the form of the scaling 
function 
\BEQ \label{R}
f_R(x) = r_0\, x^{1+a-\lambda_R/z} (x-1)^{-1-a}
\EEQ
is completely fixed ($r_0$ is a normalization constant). Eq.~(\ref{R}) has been
confirmed in several models, especially, through extensive simulations, 
in the $2D$ and $3D$ Glauber-Ising models \cite{Henk01}.

We are interested in the value of the non-equilibrium exponent $a$. In 
quite a few models, values of $a$ were obtained, see \cite{Godr02} for
a review, but no clear picture has yet emerged. 
In the exactly solved $1D$ Glauber-Ising model, 
one has $a=0$ \cite{Godr00a,Lipi00}. However, the
value of $a$ in the $2D$ and $3D$ kinetic Glauber-Ising models and the
kinetic spherical models has been debated recently. 
In the $2D$ and $3D$ Glauber-Ising model, 
analytical \cite{Bert99} and numerical \cite{Henk01} results 
indicate $a=1/2$. In the exactly solvable spherical model, one reads off 
$a=d/2-1$ for all spatial dimensions $d>2$ from the exact result for $R(t,s)$ 
\cite{Cugl95a,Zipp00,Godr00b,Pico02,Corb02a}. 

An alternative route has been followed by 
Corberi, Lippiello and Zannetti \cite{Corb01,Corb02a,Corb02b}. 
They studied the 
two-time correlation function $C(t,s)$ and the susceptibility function 
$\chi(t,s)$ in the O($n$) vector model. Separating the two-time
autocorrelation function $C(t,s) = C_{\rm st}(t-s) + C_{\rm age}(t/s)$
into a stationary part and an ageing part and similarly for the
ZFC susceptibility 
$\chi(t,s)=\chi_{\rm st}(t-s) +\chi_{\rm age}(t,s)$, where $\chi_{\rm st}$
is {\em defined} such that it satisfies the fluctuation-dissipation theorem
together with $C_{\rm eq}$, they extract an exponent
$\hat{a}$ from the scaling behaviour
$\chi_{\rm age}(t,s)=s^{-\hat{a}} \hat{\chi}(t/s)$ and claim that
$\hat{a}=a$. This was backed up by the assertion that the scaling of
$\chi_{\rm age}(t,s)$ were anomalous \cite{Corb02a}. The same
procedure was applied to simulational data of the ZFC susceptibility 
from the $2D$ and $3D$ Glauber-Ising
model. If $a_n$ is the value of $\hat{a}$ in the O($n$)-model, they propose
\cite{Corb01,Corb02a,Corb02b}
\BEQ \label{a:Corb}
a_1 = \left\{ \begin{array}{ll} (d-1)/4 & ~;~ d<3 \\ 1/2 & ~;~ d>3
\end{array}\right.
\;\; , \;\;
a_{\infty} =\left\{\begin{array}{ll} (d-2)/2 & ~;~2<d<4\\1 & ~;~ d>4
\end{array}\right.
\EEQ
for the Glauber-Ising ($n=1$) and the spherical model ($n=\infty$), 
respectively. However, since $a_{\infty}\neq d/2-1=a$ for the spherical model,
although even Corberi et {\em al.} do recover $a=d/2-1$ from $R(t,s)$, see
eq. (83) in \cite{Corb02a}, their claim of an anomalous scaling of 
$\chi(t,s)$ as raised in \cite{Corb02a,Corb02b} appears to be ill-founded. 

Our results in this letter are as follows: 
one should distinguish between those systems (called here {\it class S}) 
with a short-ranged (exponential) decay of the equilibrium 
connected spin-spin correlator 
in the ordered phase and those (called here {\it class L}) for which there 
remain long-range (algebraic) correlations in the ordered phase. 
The Glauber-Ising model in $d>1$ belongs to class S while the spherical model 
and the $2D$ XY-model are members of class L \cite{Note:Gl}. Then 
\BEQ \label{gl:a}
a = \left\{ \begin{array}{ll} 1/z & \mbox{\rm ~~; class S} \\
(d-2+\eta)/z & \mbox{\rm ~~;  class L}
\end{array} \right.
\EEQ
where $\eta$ is a well-known equilibrium critical exponent. 
Since for the Glauber-Ising model $z=2$ \cite{Bray94}, one recovers the
well-known $a=1/2$ for class S in all dimensions $d>1$. 
These results will be derived by 
comparing the scaling form of the dissipative part $\chi''$ 
of the non-equilibrium
susceptibility (see eq.~(\ref{gl:chifinal}) below) with heuristic expectations 
which for class S amount to the generally accepted 
idea that the ageing comes from the movement of the domain walls which separate 
the ordered domains \cite{Bouc00,Bray94,Bert99}. 
In addition, we shall derive a more complete scaling form for the 
thermoremanent magnetization $M_{\rm TRM}(t,s)$ in the limit of large
waiting times $s\gg 1$
\BEQ \label{gl:M-s}
M_{\rm TRM}(t,s)/h = \int_{0}^{s} \!\D u\, R(t,u) 
= s^{-a} f_M(t/s) + s^{-\lambda_R/z} g_M(t/s)
\EEQ
provided the system was initially prepared at infinite temperature. 
Eq.~(\ref{gl:M-s}) implies that the behaviour of 
$M_{\rm TRM}$ will be one of a cross-over between two 
distinct regimes \cite{Zipp00,Corb02b}. 
In practice, the cross-over region may well be large. 
If local scale invariance \cite{Henk02} holds, the scaling functions can be
found from eq.~(\ref{R}) and read explicitly 
($r_{0,1}$ are non-universal constants)
\BEQ \label{gl:Fm} 
f_M(x) = r_0\, x^{-\lambda_R/z}\, {_2F_1}\left( 1+a, \frac{\lambda_R}{z}-a;
\frac{\lambda_R}{z}-a+1; x^{-1} \right) \;\; , \;\;
g_M(x) \approx r_1\, x^{-\lambda_R/z}
\EEQ
The new and explicit forms of these scaling 
functions allow for a considerably
more precise test of the cross-over scaling of $M_{\rm TRM}$ than had been 
possible before. In particular, having fixed the parameters $r_{0,1}$ for a
given value of $x$, a precise prediction for the scaling of $M_{\rm TRM}$ for
{\em any} other value of $x$, without any free parameter, is obtained. 
We shall use this idea to perform a new type of precision test
on the scaling of $M_{\rm TRM}(t,s)$ for class S systems and in particular
test for the value of the exponent $a$.

We shall first derive the scaling forms of $\chi''$ and $M_{\rm TRM}$ 
as a function of the waiting time $s$ and
prove (\ref{gl:a}) by relating heuristic ideas on coarsening to scaling 
arguments. Then we shall describe numerical tests of (\ref{gl:M-s},\ref{gl:Fm}) 
in the spherical and Glauber-Ising models.

Consider the time-dependent response of a ferromagnetic system to a 
sinusoidal magnetic field with frequency $\omega$. The 
imaginary part of the susceptibility is given by \cite{Bouc00,Cugl02}
\BEQ
\chi''(\omega;s) = \int_{0}^{s} \!\D u\, R(s,u) \sin\left( \omega(s-u)\right)
\EEQ
which can be analysed using properties of the autoresponse function $R(s,u)$.
Here, the time difference $\tau=s-u$ plays a central role and depending on
its value an equilibrium or else an ageing behaviour is obtained. Specifically,
it can be shown \cite{Zipp00} that there is a time scale $t_p \sim s^{\zeta}$ 
such that $0<\zeta<1$ (explicitly, $\zeta=4/(d+2)$ for the spherical 
model \cite{Zipp00}) on which the breaking of the fluctuation-dissipation
relation occurs and furthermore 
$R(t,s)\simeq R_{\rm eq}(t-s)$ for $t-s\lesssim t_p$ and 
$R(t,s)=R_{\rm scal}(t,s)$ is given
by (\ref{gl:CR}) for $t_p\lesssim t-s$. For $u\approx s$ one measures the
response with respect to a change in the initial conditions and 
$R\approx R_{\rm ini}(t)\sim t^{-\lambda_R/z}$ \cite{Bray94}. Besides $t_p$, 
and following \cite{Zipp00,Cugl02}, we therefore introduce a third time scale 
$t_{\eps}$ such that $s-t_{\eps}= {\rm O}(1)$. 
Changing variables and splitting the integral into three terms, one has
\BEA
\lefteqn{\chi''(\omega;s) = \int_{0}^{s} \!\D \tau\, R(s,s-\tau) 
\sin \omega\tau }
\nonumber \\
&=& \int_{0}^{t_p} \!\D \tau\, R(s,s-\tau) \sin \omega\tau + 
\int_{t_p}^{t_{\eps}} \!\D \tau\, R(s,s-\tau) \sin \omega\tau + 
\int_{t_{\eps}}^{s} \!\D \tau\, R(s,s-\tau) \sin \omega\tau
\\
&\simeq& \int_{0}^{t_p} \!\D \tau\, R_{\rm eq}(\tau) \sin \omega\tau +
s^{-a} \int_{t_p/s}^{t_{\eps}/s} \!\D v\, 
f_R\left(\frac{1}{1-v}\right)\, \frac{\sin(\omega s v)}{(1-v)^{1+a}} + 
s^{-\lambda_R/z} \int_{t_{\eps}}^{s} \!\D \tau\, c_{0} \sin\omega\tau
\nonumber
\EEA 
where $c_0$ is a constant.   
In the third line, we used the asymptotic form of $R(s,u)$ in the three 
regimes. This is justified if for sufficiently large values of $s$, the 
cross-over between these regimes is rapid. In a given system, this can be
checked by looking for a rapid cross-over in the fluctuation-dissipation
ratio (see e.g. \cite{Bouc00,Godr02,Cugl02}) between the initial equilibrium 
and the later ageing regime, see \cite{Godr00b,Corb02a}. Now, in the 
limit $s\to\infty$ one has (i) $t_p\to\infty$, (ii) $t_p/s\to 0$ and (iii)
$t_{\eps}/s\to 1$. We then find
\BEQ \label{gl:chifinal}
\chi''(\omega;s) = \chi_1(\omega) + s^{-a} \chi_2(\omega s) + 
{\rm O}\left( s^{-\lambda_R/z} \right) 
\EEQ

For systems of class S, it is generally expected that the dynamics proceeds 
via domain growth, due to interface motion between well-ordered domains 
\cite{Bouc00,Bray94,Cugl95a,Bert99}. Therefore, the
susceptibility $\chi$ should decompose into a stationary part $\chi_{\rm eq}$
to which the spins in the bulk of the domains contribute and an ageing part
$\chi_{\rm ag}$ which comes only from the domain walls, 
with a length scale given by the typical distance $L(s)$. The 
imaginary part of the susceptibility should read \cite{Bouc00,Cugl95a,Bert99}
\BEQ \label{gl:chiIMAG}
\chi''(\omega;s) = \chi_{\rm eq}''(\omega) + L(s)^{-1}\chi_{\rm ag}''(\omega s)
\EEQ
Comparing with the general scaling form (\ref{gl:chifinal}), we see that
the first term corresponds to the equilibrium term in (\ref{gl:chiIMAG}),
the second to the ageing contribution and the third is a correction term.
{}From the scaling $L(s)\sim s^{1/z}$ of the domains one indeed recovers
the first part of (\ref{gl:a}). 

For systems of class L the long-range fluctuations turn the ordered domains
into fractal objects and we can no longer assume the existence of simple and
well-separated domain walls. Rather, we anticipate instead of 
(\ref{gl:chiIMAG}) the form
\BEQ \label{gl:chiIMAG2}
\chi''(\omega;s) = \chi_{\rm eq}''(\omega) + L(s)^{-(d-2+\eta)} 
\chi_{\rm ag}''(\omega s)
\EEQ
(up to possible logarithmic factors, which arise e.g. in the $2D$ XY model
\cite{Bert01}). 
The ageing part should come from the moving fractal domain boundaries. The
simplest way to describe the time-dependence of this is to assume that all 
length scales are measured in terms of the domain size $L(t)$. If that is so, 
the spin-spin correlator should read 
$C(\vec{r},t)\approx C_{\rm qs}(\vec{r},L(t))
\sim r^{-(d-2+\eta)} c(r/L(t))$ in terms of a quasi-stationary correlator
$C_{\rm qs}$ and where $c$ is some scaling function. The
time-dependent contribution to the susceptibility per volume $V$ 
should then come from
$\chi_{\rm qs} \sim V^{-1}\int_V \!\D\vec{r}\, C_{\rm qs}(\vec{r},L(t))\sim 
L(t)^{-(d-2+\eta)}$ and this leads to 
the desired result (\ref{gl:chiIMAG2}). The rest of the analysis then goes 
through as before and the second part of (\ref{gl:a}) follows. 
Of course, we must assume that the scaling functions $\chi_{\rm ag}''(x)$
have no singularity as $x\to\infty$. 

Eq.~(\ref{gl:a}) is confirmed for class L by all known results for simple 
ferromagnets. First, for the short-ranged spherical model either without 
\cite{Cugl95a,Zipp00,Godr00b,Godr02} or else with \cite{Pico02} initial 
long-range correlations one has $z=2$, $\eta=0$. From the exact results for 
$R(t,s)$ one recovers indeed $a=d/2-1$ as expected from (\ref{gl:a}). 
Second, for the spherical model with long-range
interactions of the form $J(\vec{r})\sim r^{-d-\sigma}$ and 
$0<\sigma<\min(d,2)$ one has $z=\sigma$, $\eta=2-\sigma$ and 
$a=d/\sigma-1$ \cite{Cann01}, as it should be. In these models, 
$b=0$ (see eq.\ (\ref{gl:CR})). 
Third, in the $2D$ XY-model one has $z=2$ \cite{Bray94} and $a=b=\eta(T)/2$ 
for both fully ordered and fully disordered initial conditions as shown 
analytically and through simulations \cite{Bert01}, 
where $\eta(T)$ depends on temperature. 
Fourth, the defining condition for class L is also satisfied 
for non-equilibrium {\em critical} dynamics. There, it is known that 
$a=b=2\beta/\nu z=(d-2+\eta)/z$, see \cite{Godr02}, 
in agreement with (\ref{gl:a}).  
Finally, for the $1D$ Glauber-Ising model ageing occurs only at the 
critical point $T_c=0$. The known exponent $\eta=1$ yields $a=0$, as it
should be \cite{Godr02}.

In the same way, we now analyse the scaling of the thermoremanent
magnetization. We stress that no use is made of eq. (\ref{gl:a}) at this 
point. Introducing the same splitting of the integral, we find
from the definition (\ref{gl:M-s}) the scaling form 
$M_{\rm TRM}(t,s)/h = \rho_0 + s^{-a} f_M(t/s) + s^{-\lambda_R/z} g_M(t/s)$
where we used the asymptotic forms of $R$ of the
three regimes. Then $f_M(x)$ is related to $f_R(x)$ \cite{Henk02} and the 
last term (and consequently $g_M(x)$, see eq.~(\ref{gl:Fm})) 
was roughly estimated from the mean-value theorem. 
The term $\rho_0$ is related to the initial
state of the system. In particular, $\rho_0=0$ for an
initially fully disordered state. While this expression for $M_{\rm TRM}$ 
depends only on the scaling
(\ref{gl:CR}), from local scale invariance as introduced and tested in 
\cite{Henk02,Henk01} we have eq.~(\ref{R}) and recover (\ref{gl:Fm}).

\begin{figure}[t]
\centerline{\psfig{figure=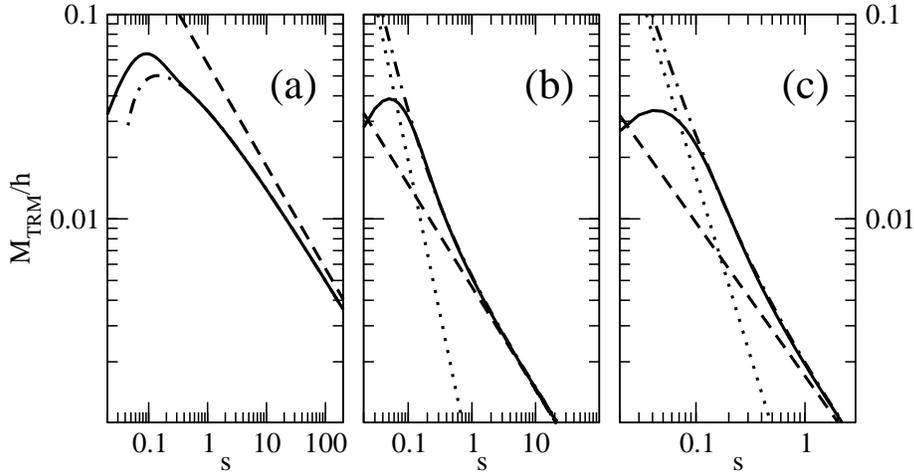,width=4.75in,clip=}}
\caption{Scaling of the thermoremanent magnetization $M_{\rm TRM}(t,s)$
(full curve) at $T=2$ as a function of the waiting time $s$ in the spherical 
model for $x=t/s=5$. (a) $d=3$ and $S_0=0$,
(b) $d=3$ and $S_0=0.5$, (c) $d=3.5$ and $S_0=0.5$. 
The dashed and dotted lines show the leading contributions
$\sim s^{-a}$  and $\sim s^{-\lambda_R/z}$, respectively, whereas the
dash-dotted lines give their sum.
\label{Bild1}}
\end{figure}

We are now ready for numerical tests of (\ref{gl:M-s}) and (\ref{gl:Fm}). 
We begin with the kinetic 
spherical model, formulated in terms of real spin variables
$S_{\vec{x}}(t)$ subject to the (mean) spherical constraint 
$\sum_{\vec{x}} \left\langle S_{\vec{x}}(t)^2 \right\rangle = {\cal N}$
where $\cal N$ is the number of sites (see \cite{Fusc02} for a careful 
discussion on this point). The kinetics is given through 
a Langevin equation 
\BEQ
\frac{\D S_{\vec{x}}(t)}{\D t} = \sum_{\vec{y}(\vec{x})}S_{\vec{y}}(t) 
-(2d+\mathfrak{z}(t))S_{\vec{x}}(t) + \eta_{\vec{x}}(t)
\EEQ
where the sum over $\vec{y}$ extends over the nearest neighbours of $\vec{x}$. 
The thermal noise is assumed to satisfy
$\langle\eta_{\vec{x}}(t) \eta_{\vec{y}}(t')\rangle =
2 T\,\delta_{\vec{x},\vec{y}}  \delta (t-t')$. From the
spherical constraint the Lagrange multiplier
$\mathfrak{z}(t)$ can be found through
the solution of a Volterra integral equation, and two-time
correlators and response functions are readily found, 
see \cite{Cugl95a,Godr00b,Paes03} for details. 

In figure~\ref{Bild1}
we show the thermoremanent magnetization $M_{\rm TRM}$
obtained from numerically integrating the Langevin equation, 
at a temperature $T<T_c$ (recall that for $d=3$, $T_c\simeq 3.96$ and 
for $d=3.5$, $T_c\simeq 5.27$). 
As an initial state, we used uncorrelated spins with either a mean
magnetization $S_0=0$ (figure~\ref{Bild1}a) 
or else $S_0=0.5$ (figure~\ref{Bild1}b and \ref{Bild1}c).
Then the exponents $z=2$, $a=d/2-1$ and $\lambda_R=d/2$ for $S_0=0$ 
or $\lambda_R=d$ for $S_0=0.5$, respectively, are expected \cite{Pico02}. 
The leading term $\sim s^{-a}$ and the sum of the two leading contributions to
$M_{\rm TRM}$  according to eq.~(\ref{gl:M-s}) are shown and are 
compared with the numerical solution of the Langevin equation \cite{Note:r01}. 
For $S_0=0.5$, the cross-over between the two regimes is evident. On the
other hand, for $S_0=0$, we observe that although the slope of 
$\ln M_{\rm TRM}$ versus $\ln s$ appears
to be fairly constant, the second term expected from eq.~(\ref{gl:M-s})
produces a sizable correction. We shall
find a similar behaviour for the $2D$ Glauber-Ising model below. 
Finally, we see that down to values of $s$
as small as $s\sim 1$, the time-dependence of $M_{\rm TRM}$ is well described
by eq.~(\ref{gl:M-s}). We observed a similar behaviour, and in quantitative 
agreement with the scaling functions (\ref{gl:Fm}), 
for other values of $x$.

\begin{figure}[t]
\centerline{\epsfxsize=3.75in\epsfbox
{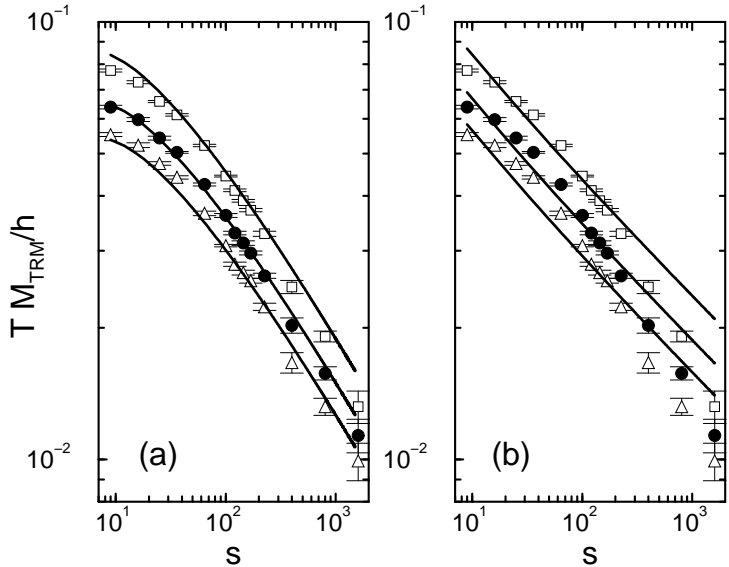}
}
\caption{Scaling of the thermoremanent magnetization $M_{\rm TRM}(t,s)$ as
a function of $s$ for several values of $x=t/s$ (squares: $x=5$, circles:
$x=7$, triangles: $x=9$), as obtained in the $2D$
Glauber-Ising model at $T=1.5$. The black curves are calculated from 
\protect{equations~(\ref{gl:M-s}) and (\ref{gl:Fm})} with 
$\lambda_R=1.26$, see \cite{Fish88,Henk01};
(a) $a=1/2$ and (b) $a=1/4$. 
\label{Bild2}}
\end{figure}

We now come to our main objective: deciding between $a=1/2$ and $a=1/4$ through
a test of (\ref{gl:M-s}) in the $2D$ Glauber-Ising model. 
This model describes the evolution of the spin variables $s_{j}=\pm 1$ on the 
sites $j$ of a hypercubic lattice, realized through heat-bath
Glauber dynamics. We use a temperature $T=1.5<T_c$ and a spatially random
field $h=\pm 0.05$ and measure
$M_{\rm TRM}(t,s)$ in the standard way \cite{Barr98}. 
Starting from a totally disordered state, we measure the evolution of 
$M_{\rm TRM}(t,s)$ as a function of $s$ with $x=t/s$ fixed. 
The non-universal parameters $r_{0,1}$ were fixed using the data for $x=7$
for either $a=1/2$ ($r_0 = 1.76$, $r_1= - 1.84$) or
$a=1/4$ ($r_0 = 0.22$, $r_1 = 0.09$), respectively. 
Since $r_{0,1}$ are independent of $x$, we therefore obtain a prediction on how
$M_{\rm TRM}(xs,s)$ should scale for any other value of $x$. This is
shown in figure~\ref{Bild2}.  
While for $a=1/2$, the Monte Carlo data nicely agree with the 
cross-over scaling as predicted from eqs.~(\ref{gl:M-s},\ref{gl:Fm}), 
it is also clear from
figure~\ref{Bild2}b that the hypothesis $a=1/4$ \cite{Corb02a,Corb02b} is 
incompatible with the data. 
The same conclusion, namely the invalidity of the value $a=1/4$
in the $2D$ Glauber-Ising model, has also been reached using simulational
data at $T=0$ with waiting times up to $s=5600$ \cite{Henk03}.

Summarizing, we have reanalysed the scaling of the magnetic linear response
in simple ferromagnets undergoing phase-ordering kinetics. 
The cross-over scaling of the thermoremanent magnetiztion eq.~(\ref{gl:M-s}) 
(and also the dynamic susceptibility) is predicted exactly for any dynamic 
universality class from local scale invariance in terms of the 
universal exponents $\lambda_R/z$ and $a$. We have confirmed 
eqs.~(\ref{gl:a}) and (\ref{gl:M-s},\ref{gl:Fm}) 
in the Glauber-Ising and spherical models. 
This provides evidence against recent suggestions of an anomalous scaling of
the dynamic suceptibilities in phase-ordering kinetics.

\acknowledgments
M. Pae{\ss}ens is grateful to the Deutscher Akademischer Austauschdienst 
(DAAD) for financial support. 
This work was supported by the Bayerisch-Franz\"osisches Hochschulzentrum 
(BFHZ) and by CINES Montpellier (projet pmn2095).



\begin{thebibliography}{999}
\bibitem{Note:Gl} The kinetics of the $1D$ Glauber-Ising model is very 
peculiar since its critical temperature is $T_c=0$. It appears to fit into 
class L rather than class S.
\end{thebibliography}
\end{document}